\documentclass[10pt,b5paper,twoside,onecolumn]{scrartcl}
\usepackage[cp1250]{inputenc}
\usepackage{color,fancyvrb}
\usepackage{graphicx}
\usepackage{hyperref}
\usepackage{fancyhdr}
\usepackage{setspace}
\usepackage{float}
\usepackage{pdfpages}
\usepackage{verbatim}
\usepackage{lmodern} 
\usepackage{amsthm}
\usepackage{amssymb}
\usepackage{multicol}
\usepackage{booktabs}
\usepackage{mathpazo}
\usepackage[titles]{tocloft}
\begin{document}

\pagestyle{fancyplain}
\fancyhf{}
\fancyhead[LE]{\textit{A Probabilistic Ant-based Heuristic for the Longest Simple Cycle Problem in Complex Networks}}
\fancyhead[RO]{\textit{D. Chalupa, P. Balaghan, K. A. Hawick}}
\fancyfoot[C]{\thepage}
\fancypagestyle{plain}
{
	\fancyhf{} 
	\renewcommand{\headrulewidth}{0pt} 
	\renewcommand{\footrulewidth}{0pt}
}

\thispagestyle{empty}

\begin{center}\textbf{\LARGE\sffamily\noindent
A Probabilistic Ant-based Heuristic for the Longest Simple Cycle Problem in Complex Networks
}\end{center}

\begin{center}{\large\sffamily\noindent David Chalupa$^a$, Phininder Balaghan$^b$, Ken A. Hawick$^b$}\end{center}
\begin{center}
{
\noindent
$^a$~Operations Research Group\\
Department of Materials and Production\\
Aalborg University\\
Fibigerstr\ae de 16, Aalborg 9220, Denmark\\
Email: \texttt{dc@m-tech.aau.dk}\\
$^b$~School of Engineering and Computer Science\\
University of Hull\\
Cottingham Road\\
Hull HU6 7RX, United Kingdom\\
Email: \texttt{\{p.balaghan,k.a.hawick\}@hull.ac.uk}

}
\end{center}

\vspace{30pt}

\paragraph{Abstract.} We propose a new probabilistic ant-based heuristic (ANTH-LS) for the longest simple cycle problem. This NP-hard problem has numerous real-world applications in complex networks, including efficient construction of graph layouts, analysis of social networks or bioinformatics.
Our algorithm is based on reinforcing the probability of traversing the edges, which have not been present in the long cycles found so far. Experimental results are presented for a set of social networks, protein-protein interation networks, network science graphs and DIMACS graphs. For 6 out of our 22 real-world network test instances, ANTH-LS has obtained an improvement on the longest cycle ever found.


\section{Introduction}\label{sec:intro}

In many real-world network applications, the problem of finding the longest simple cycle is of a high interest. These applications include suitable plane separation in graph drawing \cite{TamassiaHandbookOfGraphDrawing}, social network analysis \cite{CSI-0055}, or analysis of metabolic pathways in bioinformatics \cite{BeckerGraphLayoutMetabolicPathways}, reconstructed from protein-protein interactions \cite{CohenIntroductionBioinformatics}.

The problem is NP-hard as a consequence of the NP-hardness of Hamiltonian cycle problem \cite{KarpReducibilityAmongCombinatorialProblems}. However, it is somewhat more intriguing than most NP-hard problem to solve computationally. This is because the currently available formulations model the problem as a sequence of integer linear programming (ILP) instances, sampling the longest cycle from a fixed vertex \cite{CSI-0055,DixonAlgorithmLongestCycle}. Each of these instances can be computationally hard to solve.

\begin{figure*}
\begin{center}
\includegraphics[scale=0.135]{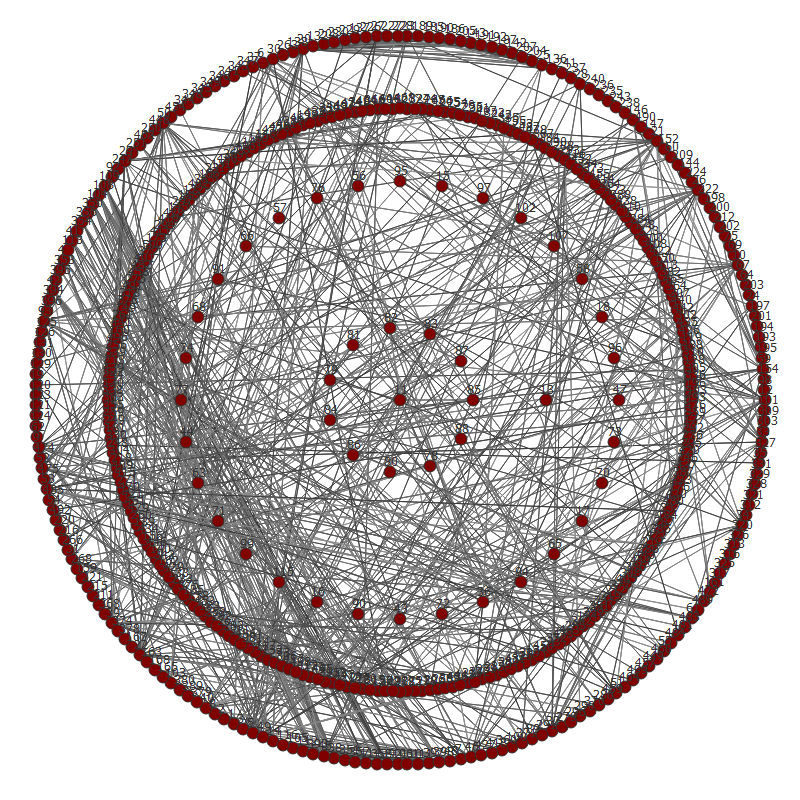}
\includegraphics[scale=0.135]{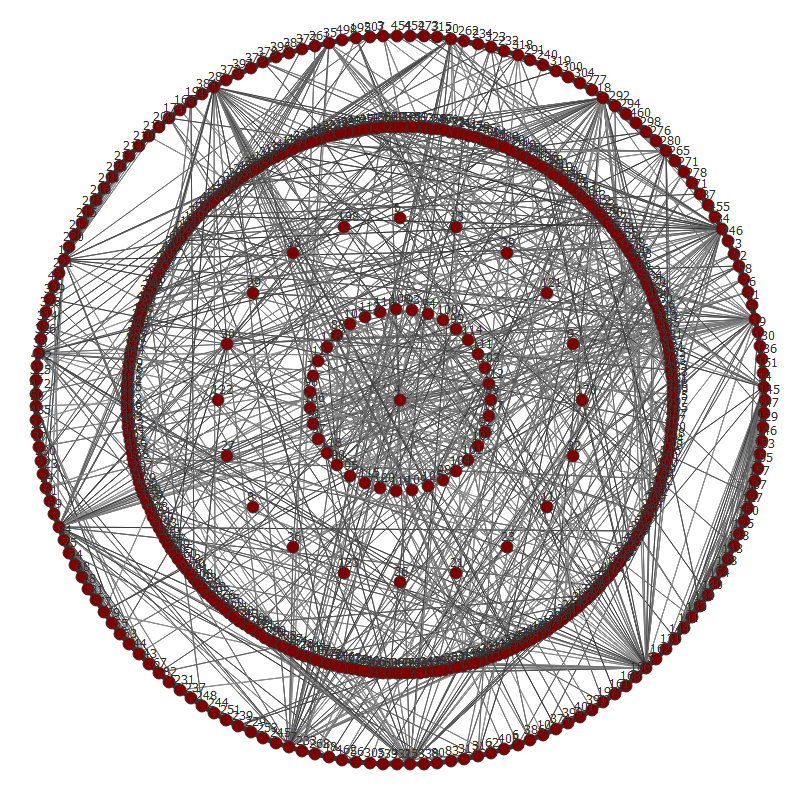}
\includegraphics[scale=0.135]{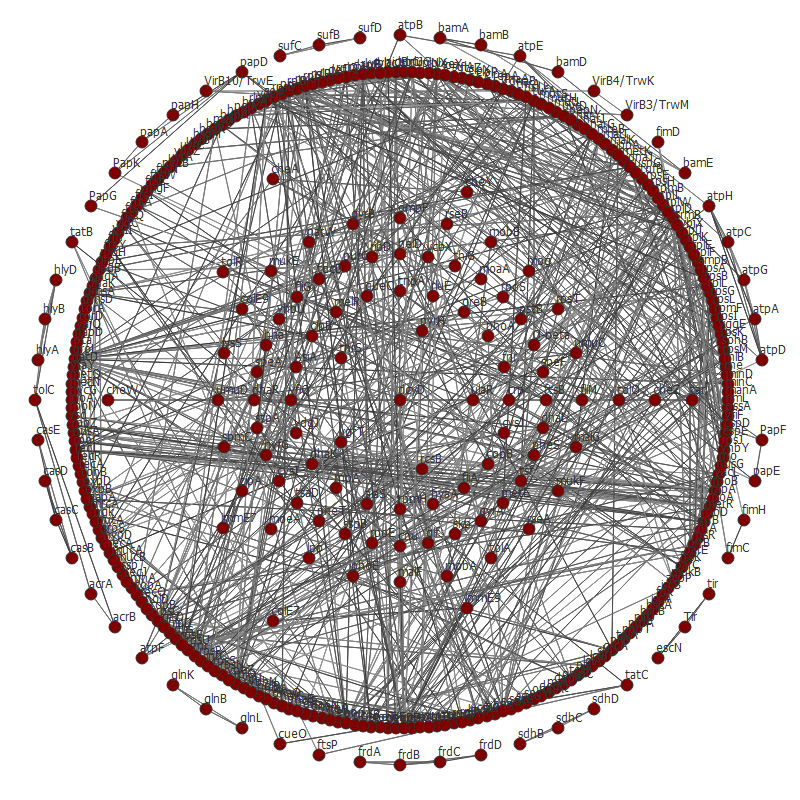}\\
(a) \hspace{93pt} (b) \hspace{93pt} (c)\\
\includegraphics[scale=0.135]{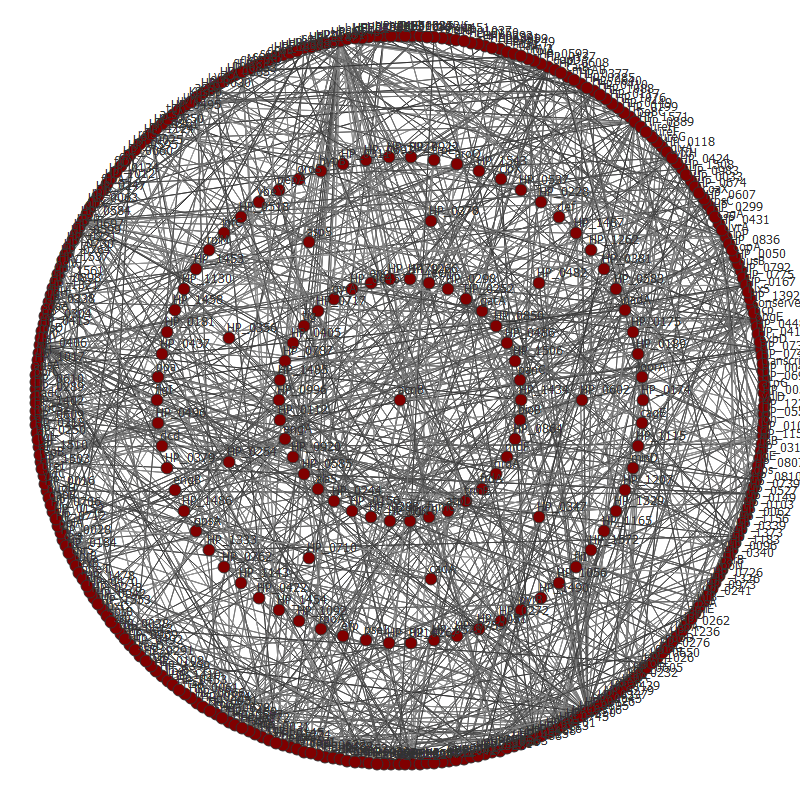}
\includegraphics[scale=0.135]{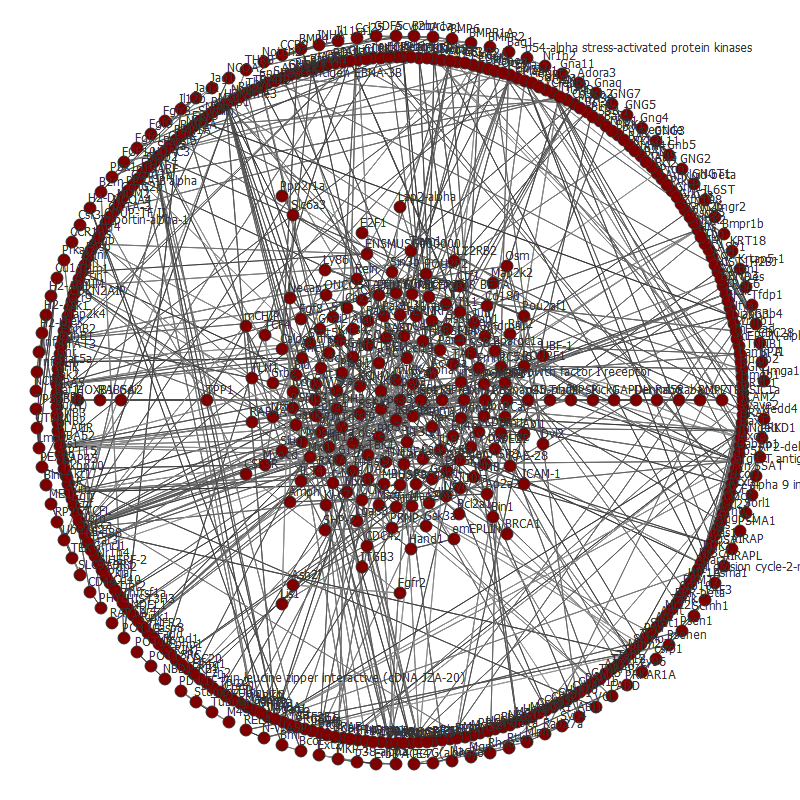}
\includegraphics[scale=0.135]{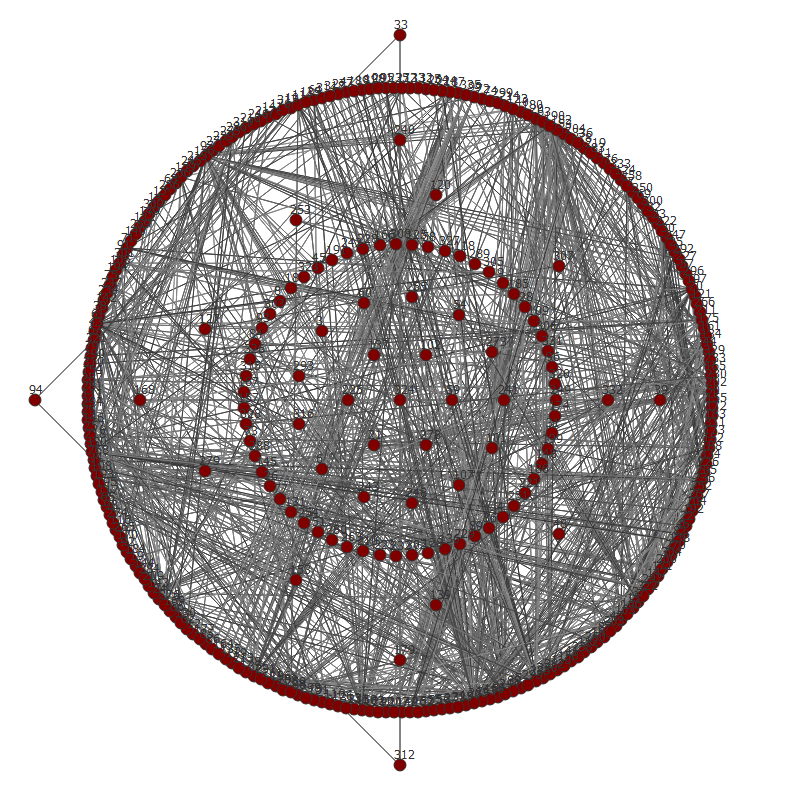}\\
(d) \hspace{93pt} (e) \hspace{93pt} (f)
\end{center}
\caption{Illustrations of the 6 cycles longest cycles obtained by the ANTH-LS algorithm, improving on the best result ever found. The cycles found are represented by the outer circles within the drawings. These represent cycles of lengths (a) $215$ for $gplus\_500$, (b) $167$ for $pokec\_500$, (c) $246$ for $Ecoli20160114CR$, (d) $311$ for $Hpylo20160114$, (e) $278$ for $Mmusc20160114CR$ and (f) $242$ for $homer$. Drawings (c)-(f) depict the cycles within subgraphs obtained by pruning the leaves.}
\label{fig:teaser}
\end{figure*}

\paragraph{Contributions.}
In this paper we explore the use of an ant colony optimisation framework \cite{DorigoAntColonyTheory} to solve the longest simple cycle problem. We propose a new probabilitistic ant-based heuristic (ANTH-LS) to solve the problem in real-world complex networks. Somewhat surprisingly, we discover that reinforcing the probability of traversing the edges, that have not been used in the previous long cycles sampled, leads to a highly successful strategy to solve the problem in large-scale. Conversely, the other edges are traversed with decreased probability.

The ANTH-LS algorithm uses the idea of ant colony optimisation in a way somewhat similar to tabu search, effectively decreasing the probability of reattempting previously attempted moves. Each cycle is improved by a subroutine using four perturbation operators that have been previously used in multi-start local search \cite{CSI-0055}. Our new algorithm outperforms both the multi-start local search heuristic currently available for the problem, as well as an ILP-based solver with a generous time limit of one hour per vertex. Experimental results are presented for a set of 22 real-world complex networks, including social network samples, network science data, protein-protein interaction networks and coappearance networks for literary classics. For 6 out of these 22 instances, new longest cycles cycles ever found have been discovered by the ANTH-LS algorithm. These are depicted in Figure 1.

The structure of the paper is as follows. Section 2 features an overview of the problem and related work. In Section 3, we propose our new ANTH-LS algorithm for the longest simple cycle problem. In Section 4, we present the experimental results and provide a brief discussion. Last but not least, Section 5 summarises the work and identifies open problems.

\section{The Longest Simple Cycle Problem}\label{sec:lc}

Identification of long cycles in complex networks has previously attracted a considerable attention in statistical mechanics \cite{MarinariFindingLongCycles}. From the algorithmic perspective, enumerative algorithms have been explored for some time in finding all cycles in a complex network \cite{JohnsonFindingAllElementaryCircuits,TarjanEnumerationOfElementaryCircuits}. However, these algorithms unsuprisingly lack in scalability and can only be used up to graphs of certain sizes, with the notable example of a $7 \times 7$ grid as a borderline case in practice \cite{CSTN-013}.

The idea of combining depth-first search (DFS) with dynamic programming has been previously used to construct an approximation algorithm for the problem \cite{BodlaenderMinorTestsDepthFirstSearch}. However, approximation algorithms seem to be mainly of theoretical interest in this problem so far.

A classical integer linear programming (ILP) formulation of the problem is due to Dixon and Goodman, who also proposed a branch-bound algorithm to solve the problem \cite{DixonAlgorithmLongestCycle}. This formulation used the idea of introducing a ``dummy'' vertex to the network and searching for the longest path from a fixed vertex to this new vertex. In their study, the problem was considered for graphs with up to $40$ vertices.

In our more recent study, we proposed a new ILP formulation based on flow constraints \cite{CSI-0055}, inspired by a similar formulation of the travelling salesperson problem (TSP) \cite{TSPFlowBased}. We conducted experiments using a branch-and-cut mixed-integer linear programming solver CBC from COIN-OR package \cite{BonamiAlgorithmicMixedIntegerPrograms,LinderothMilp}, enhanced by our own data mining and aggregation pipeline. We discovered that CBC is able to solve the problem much more efficiently when it is formulated using the flow-based constraints. We also introduced a multi-start local search (MSLS) heuristic that combined repeated sampling of promising cycles using DFS with four perturbation operations used to enhance the initial cycle.

The idea to use ant-based algorithms to solve tour-based problems has been explored previously, as ant-based algorithms usually have a good performance in ``pathfinding'' problems \cite{DorigoAntColonyTheory}. Notably, an ant-based algorithm has been used to solve the related Hamiltonian cycle problem \cite{WagnerAntInspiredHeuristicHamiltonianGraphs}. However, the idea to use this framework to find long cycles has not been explored yet, to the best of our knowledge. 

\section{An Ant-based Heuristic (ANTH-LS) for the Longest Simple Cycle Problem}\label{sec:acolc}

In DFS, each potential transition from a vertex to another vertex occurs with the same probability. This may be desirable in random sampling of potential solutions with a considerable level of diversity. However, once several long cycles are identified, it is interesting to incorporate this information into a probabilistic model of the problem and its instance.

Our ant-based algorithm ANTH-LS will expand on the DFS framework by assigning a pheromone value $\tau_e$ to each edge $e \in E$ of the graph. This value will influence the probability of the edge being traversed while searching for the longest cycle. The higher value of $\tau_e$ is, the higher will also be the probability that $\tau_e$ will be chosen as the next edge to be traversed in the search process.

In our preliminary experiments, we realised that the model does not seem to benefit from reinforcing the probability of traversing the edges in the long cycles previously found. To our surprise, applying the idea inversely by reinforcing the previously unexplored edges, slowly leads the ant-based framework to a bias towards longer cycles. Our ANTH-LS algorithm will therefore iteratively reinforce the probability of traversing the edges, which have not been traversed in the previous iteration. On the other hand, it will penalise the edges, which have been traversed in the previous iteration.

\begin{table}
\begin{center}
Algorithm 1. A Probabilistic Construction Procedure for Long Cycles in ANTH-LS Algorithm Using DFS (partly adapted \cite{CSI-0055})\vspace{5pt}\\
\begin{tabular}{l|l}
& Input: graph $G = [V,E]$, pheromone values $\tau_e$ for each $e \in E$\\
& Output: cycle $C = [v_{c_1}, v_{c_2}, ..., v_{c_k}]$\\\hline
1  & $C = []$, $l = 0$\\
2  & for each $v_{central} \in V$\\
3  & ~~~~$\forall v \in V$ let $d(v) = 0$, $p(v) = 0$\\
4  & ~~~~$S = [v_{central}]$\\
5  & ~~~~while $S$ contains at least one vertex\\
6  & ~~~~~~~~$v = pop(S)$\\
7  & ~~~~~~~~if $p(v) = 0$\\
8  & ~~~~~~~~~~~~$d_c = d(v)$, $p(v) = 1$\\
9  & ~~~~~~~~~~~~choose $w$ such that $\{v,w\} \in E$ with\\
	 & ~~~~~~~~~~~~probability proportional to $\tau_{\{v,w\}}$\\
10 & ~~~~~~~~~~~~~~~~if $p(w) \neq 0 \wedge w \neq v_{central} \wedge d(w) \leq d_c + 1$\\
11 & ~~~~~~~~~~~~~~~~~~~~remove $w$ from $S$ if it currently is in $S$\\
12 & ~~~~~~~~~~~~~~~~~~~~$S = push(w)$\\
13 & ~~~~~~~~~~~~~~~~~~~~$d(w) = d_c+1$, $parent(w) = v$\\
14 & ~~~~~~~~~~~~~~~~if $w = v_{central} \wedge d_c \geq 2 \wedge l < d_c + 1$\\
15 & ~~~~~~~~~~~~~~~~~~~~$l = d_c + 1$\\
16 & ~~~~~~~~~~~~~~~~~~~~use array $parent$ to trace the current\\
   & ~~~~~~~~~~~~~~~~~~~~longest cycle $C$ to $v_{central}$\\
17 & return $C$\\
\end{tabular}
\end{center}
\end{table}

Algorithm 1 presents the probabilistic construction procedure used to construct a long cycle, based on the current pheromone values. The pseudocode is largely adapted from the previous DFS algorithm and is based on this framework \cite{CSI-0055}. However, Algorithm 1 has the pheromone values $\tau_e$ as input parameters. It samples long cycles iteratively for each vertex and uses a queue $S$ to record the vertices to visit. The core operation is performed in step 9, where the next vertex is chosen to be added to $S$, and visited in the next step of DFS. This is performed by selecting the next vertex $w$ to succeed the current vertex $v$ with probability proportional to the pheromone value on the corresponding edge. Using a simplified notation, this is determined by:

\begin{equation}
Prob(v \rightarrow w) = \frac{\tau_{\{v,w\}}}{\sum_{w': \{v,w'\} \in E} \tau_{\{v,w'\}}}.
\end{equation}

\begin{figure*}
\begin{center}
\includegraphics[width=0.9\hsize]{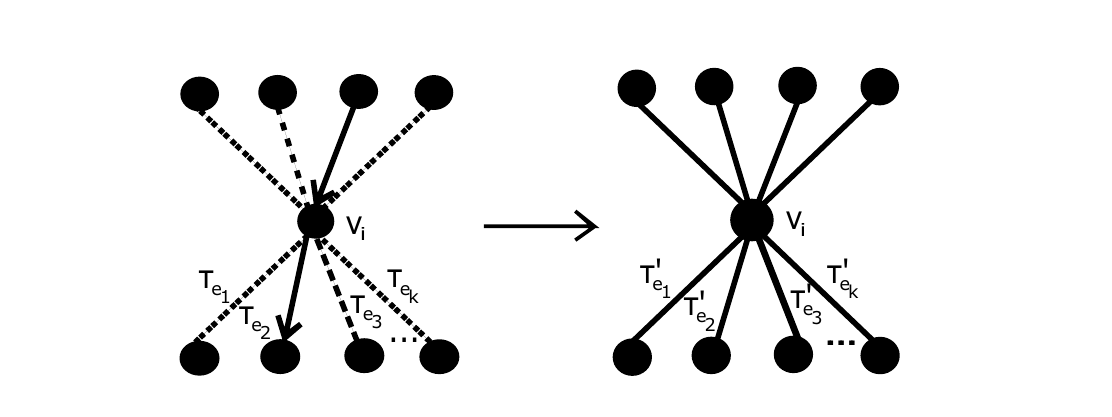}
\end{center}
\caption{An illustration of the traversals possible in the construction procedure used within the ANTH-LS algorithm. For the current vertex $v_i$, edge $e_2$ is traversed as an example. As a consequence, the new pheromone value $\tau'_{e_2} = \rho\tau_{e_2}$, while $\tau_{e_j} + \frac{1}{10-f(A_{best})+f^{*}}$ for $j \neq 2$, $1 \leq j \leq k$.}
\label{fig:update}
\end{figure*}

\noindent
In the following, we will describe the ANTH-LS algorithm simply as sequence of enumerated steps:

\begin{enumerate}
	\item For each $e \in E$, set $\tau_e = \tau_0$, i.e. deposit $\tau_0$ units of pheromone on each edge $e \in E$. We will use $\tau_0 = 10$.
	
	\item Use Algorithm 1 with the probabilistic model determined by the phero\-mone values $\tau$ to construct $A = 5$ long cycles.
	
	\item For each of the $A$ long cycles, choose local search LS-III with probability $1/2$ or LS-IV otherwise (see below for details on the local search algorithms), and improve the cycle by the chosen local search approach.
	
	\item Take the longest of these cycles $A_{best}$ of length $f(A_{best})$. Let $f^*$ be the length of the longest cycle found so far. Then, each value $\tau_e$ for $e \in E(A_{best})$, where $E(A_{best})$ is a the set of edges in the cycle $A_{best}$, will be updated as follows:
	\begin{equation}
	\tau_e = \rho \tau_e,
	\end{equation}
	and for $e \notin E(A_{best})$, the formula will be similar to a formula used in ant-based algorithms for the minimum dominating set problem \cite{PotluriTwoMetaheuristicDominatingSet} and the leaf-constrained minimum spanning tree problem \cite{SinghNewMetaheuristicLeafConstraintedMST}:
	\begin{equation}
	\tau_e = \tau_e + \frac{1}{10-f(A_{best})+f^{*}},
	\end{equation}
		where $\rho = 0.95$ is the pheromone evaporation rate. These values were adapted from value, which have been used in ant-based algorithms to solve the dominating set problem \cite{PotluriTwoMetaheuristicDominatingSet}. Note that no pheromone value is allowed to be lower than $\tau_{min} = 0.01$, to avoid premature convergence.
			
	\item The algorithm is stopped whenever $g_{conv} = 10$ consecutive generations consisted solely of cycles of the same length, indicating convergence of the algorithm, or if a maximum of $10000$ generations overall have been reached.
\end{enumerate}

\paragraph{Local search algorithms} LS-III and LS-IV are both based on a repeated application of three or four perturbation operators. These operators are depicted in Figure 3 and have previously been used within MSLS \cite{CSI-0055}. LS-III uses operators (a)-(c), while LS-IV uses all four operators depicted. Operators (a) and (c) represent the improvement operators, enlarging the cycle by substituting an edge with a ``diversion'' path of length $2$  or $3$, respectively. Informally, we refer to them also as the triangular and rectangular operators. The remaining two operators are used to explore plateaus of cycles with the same length. Operator (b) substitutes a path of length $2$ with an alternative path of length $2$. Operator (d) performs an equivalent subtitution for paths of length $3$.

LS-III and LS-IV simply repeatedly apply operators (a)-(c) or all operators to each edge or a subpath in the current cycle. Improvement operators take precedence. The local search process is terminated if no operator can be applied to the current solution to obtain at least as long cycle as the current one, or $i_{stag} = 100$ iterations without enlarging the current cycle are reached.

This local search subroutine represents an intensification component of the ANTH-LS algorithm. It potentially improves a candidate long cycle to a local optimum. In contrast to this, we will see that construction procedure serves in fact more as a diversification procedure, exploring different regions of the search space.

\begin{figure}[t]
\begin{center}
\includegraphics[width=0.24\hsize]{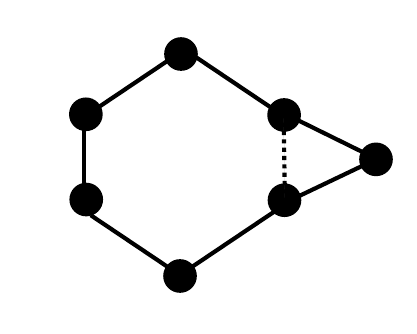}
\includegraphics[width=0.24\hsize]{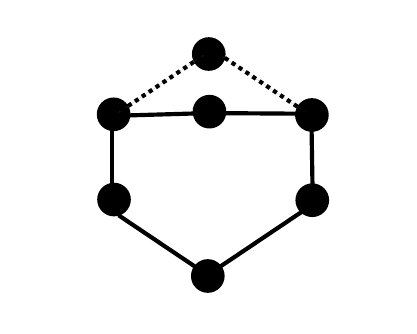}
\includegraphics[width=0.24\hsize]{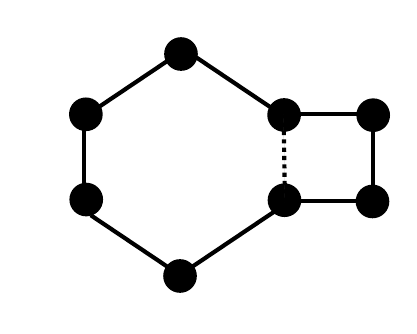}
\includegraphics[width=0.24\hsize]{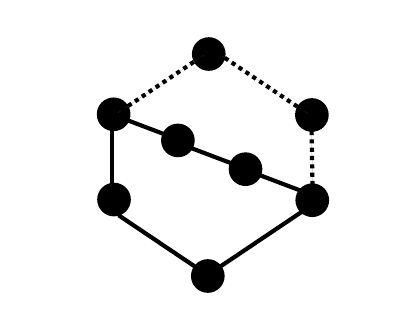}\\
(a) \hspace{72pt} (b) \hspace{62pt} (c) \hspace{72pt} (d)
\label{fig:operators}
\end{center}
\caption{The four perturbation operators used originally as intensification operators within MSLS \cite{CSI-0055} and also used in this paper within the ANTH-LS algorithm. Operators (a) and (b) are the improvement operators, while operators (c) and (d) are the plateau exploration operators. LS-III uses operators (a)-(c), while LS-IV uses all four perturbation operators.}
\label{fig:perturbations}
\end{figure}

\section{Experimental Results}\label{sec:results}

In this section, we present the experimental results of our ANTH-LS algorithm, as well as their comparison to the best results obtained by the multi-start local search (MSLS) and the ILP-based approach using CBC.

The ANTH-LS algorithm was used in a configuration with $A = 5$ individuals, pheromone evaporation rate $\rho = 0.95$ and minimum pheromone value $\tau_{min} = 0.01$. ANTH-LS was run 10 times and each run was stopped after $50$ consecutive iterations with all individuals of the same cycle length. Results of the branch-and-cut solver CBC from the COIN-OR package \cite{BonamiAlgorithmicMixedIntegerPrograms,LinderothMilp} have been taken from our previous study, as well as the results obtained by the variants of multi-start local search (MSLS) \cite{CSI-0055}. 

The ANTH-LS algorithm has been implemented in C++ using the Qt framework. The experiments were run on a Apple Mac Pro machine with OS X Sierra, 3.5GHz 6-Core Intel Xeon E5 CPU, and with 16 GB 1866 MHz DDR3 RAM. This is the same configuration that was used in the previous work, used for a comparison of the results obtained by the algorithm proposed in this paper.

For experimental evaluation, we used the same set of instances as in our previous study \cite{CSI-0055}. These instances consist of several social network samples, various graphs Newman's network science data repository, as well as protein-protein interaction networks from the UCLA database of interacting proteins \cite{SalwinskiDatabaseOfInteractingProteins,XenariosDatabaseOfInteractingProteins2001,XenariosDatabaseOfInteractingProteins,XenariosDatabaseOfInteractingProteins2002}, and coappearance networks for classical literary characters from DIMACS graphs \cite{JohnsonDimacs}. Social network $soc52$ is a simple graph representing the neighourhood of a single social network user. Instances of types $gplus$ and $pokec$ represent samples from social networks Google+ and Pokec with corresponding sizes.

Graph $adjnoun$ is an adjective-noun adjacency network, $football$ is an American college football network, $lesmis$ represents a coappearance network for Les Miserables, $netscience$ is a graph of network science collaborations, $zachary$ is a network of friendships within a karate club, $celegansneural$ is a neural network for Caenorhabditis elegans species, $dolphins$ is network of interactions of bottlenose dolphins, and $polbooks$ is a network of political books.

Protein-protein interaction networks include \textit{Celeg20160114CR} for Caenor\-habditis elegans, \textit{Dmela20160114CR} for Drosophi\-la melanogaster, \linebreak \textit{Ecoli20160114CR} for Escherichia coli, \textit{Hpylo20160114} for Helicobacter pylori, \textit{Hsapi20160114HT} is a human protein-protein interaction network and network \textit{Mmusc20160114CR} is for Mus musculus.

Coappearance network \textit{anna} is for Anna Karenina, $david$ is for David Copperfield, $homer$ is for Iliad and Odyssey and $huck$ is for Huckleberry Finn.

An interested reader may refer to the previous work for more details on the instances \cite{CSI-0055}, which we omit in this paper for shortness.

All instances were first preprocessed by pruning their leaves iteratively, since vertices in a cycle need to have a degree at least $2$. Such a simple preprocessing routine can reduce the instance size significantly.

\begin{table*}
{\scriptsize
\begin{center}
\caption{Computational results obtained for the ANTH-LS algorithm, including the length of the longest cycle found, the success rate among the $10$ independent runs, the average number of generations and the average CPU time required. For each instance, we also report the previously best known cycle, as well as the optima for the instances, for which these are known.}
\begin{tabular}{l l l l l l l l}
\toprule
graph						& optimum			& previously	&	\multicolumn{4}{c}{ANTH-LS} \\
								&							& best				&	cycle				& success		& average			& CPU  \\
								&							& known				&	length			& rate			& generations	& time \\
\midrule
\multicolumn{7}{c}{\textit{Social networks}} \\
\textit{gplus\_200}				& 	70			&	70		&	70		&	10 / 10		&	374	 &	 13 s\\
\textit{gplus\_500}				& 				& 	202	 & 215			&	3 / 10			&	1001	 & 263	s\\
\textit{pokec\_500}				& 	&		163			&	167		&		9 / 10		&	1644	&	490	s	\\
\textit{soc\_52}					&		51	& 51			&	51		&		10/ 10		&	351	&	7 s	\\
\midrule
\multicolumn{7}{c}{\textit{Graphs from Newman's network data repository}} \\
\textit{adjnoun}	 \cite{NewmanFindingCommnityStructureUsingTheEigenvectors}
						&		101	& 101 &	101		&		10 / 10		&	1232	 & 49 s	\\
\textit{football} \cite{GirvanCommunityStructureSocialBiologicalNetworks}
						& 		115	& 115	&	115		&		10 / 10		&	2072	& 272 s	\\
\textit{lesmis} \cite{KnuthStanfordGraphBase}
						& 		49	& 49	&	49		&	10	/ 10		&	557	& 13 s	\\
\textit{netscience} \cite{NewmanFindingCommnityStructureUsingTheEigenvectors}
						& 		&	108	&	108		&	9	/ 10	&	327		& 98 s 	\\
\textit{zachary}	 \cite{ZacharyConflictAndFissionInSmallGroups}
						&		20 & 20		&	20		&		10 / 10		&	139	& $<$1 s	\\
\textit{celegansneural} \cite{WattsStrogatzCollectiveDynamicsSmallWorldNetworks}
						& 	280 &	280 &	280		&		10 / 10	&	4228		& 385 s	\\
\textit{dolphins} \cite{LusseauBottlenoiseDolphins}
						& 		53	& 53			&	53		&		10	/ 10	&	570	& 2	s \\
\textit{polbooks*}				&		105	& 105	&	105		&		10 / 10		&	1273 	&	52 s	\\
\midrule
\multicolumn{7}{c}{\textit{Protein-protein interactions from UCLA database of interacting proteins} \cite{SalwinskiDatabaseOfInteractingProteins,XenariosDatabaseOfInteractingProteins2001,XenariosDatabaseOfInteractingProteins,XenariosDatabaseOfInteractingProteins2002}} \\
\textit{Celeg20160114CR}		&	6 & 6			&	6		&		10/ 10	&	9	& $<$1 s	\\
\textit{Dmela20160114CR}		&	 14 &	14	&	14		&			10 / 10	&	12	&	$<$1 s			\\
\textit{Ecoli20160114CR}			& 		& 242	&	246	&		2 / 10		&	1320	&	480 s	\\
\textit{Hpylo20160114}			& 	& 291		&	311		&		1 / 10		&	2083 	& 1345 s	\\
\textit{Hsapi20160114HT}		&		64		& 64	&	64		&		10 / 10		&	254	& 3 s	\\
\textit{Mmusc20160114CR}			& 	& 256	&	278		&		1 / 10		&	1617	&526 s	\\
\midrule
\multicolumn{7}{c}{\textit{DIMACS graphs} \cite{JohnsonDimacs}} \\
\textit{anna}					&		& 79 &	79		&		10 / 10		&	656 &	36 s\\
\textit{david}					&		72	&	72		&	72		&		10 / 10	&	637 & 21 s	\\
\textit{homer}					&		&	 223 &	242		&			6 / 10	&	1483 	& 999 s	\\
\textit{huck}					& 		48 & 48	&	48		&		10 / 10	&	305	&	5 s	\\
\bottomrule
\end{tabular}
\end{center}
\label{tab:resultsheuristic}
* Network \textit{polbooks} has not been published in a past paper. It is available from Newman's network data repository:\\
\texttt{http://www-personal.umich.edu/$\sim$mejn/netdata/}
}
\end{table*}

\begin{table*}
{\scriptsize
\begin{center}
\caption{Comparison of the longest cycles found by the ANTH-LS algorithm to the other approaches. CBC represents the results obtained by the ILP-based solver with $1$ hour time limit per vertex, and MSLS algorithms represent multi-start local search with $10000$ or $100000$ restarts, with the use of 3 or 4 perturbation operators (indicated by III or IV suffices). The lengths of the improved longest cycles ever found are highlighted in bold.}
\begin{tabular}{l l l l l l l l}
\toprule
graph						& optimum			& previously	&	CBC					& MSLS-				& MSLS-				& MSLS- 	& ANTH-LS \\
								&							& best				&	(ILP)*			& 10000-			& 10000-			& 100000-	& \\
								&							& known				&							& III					& IV					& III 		& \\
\midrule
\multicolumn{7}{c}{\textit{Social networks}} \\
\textit{gplus\_200}				& 	70			&	70		& 70		& 67		 	& 66 		 	& 68 	 & 70	\\
\textit{gplus\_500}				& 					& 	202	 & 202		& 186		 	& 186	 		& 192		&	\textbf{215} \\		
\textit{pokec\_500}				& 	&		163			& 	163  & 155 	& 151				& 159	&	\textbf{167}		\\		
\textit{soc\_52}					&		51	& 51			& 51		& 51 		& 51		 		& 51 	&		51		\\
\midrule
\multicolumn{7}{c}{\textit{Graphs from Newman's network data repository}} \\
\textit{adjnoun}	 \cite{NewmanFindingCommnityStructureUsingTheEigenvectors}
						&		101	& 101 & 101	& 91		 	& 92		 			& 93				&	101\\
\textit{football} \cite{GirvanCommunityStructureSocialBiologicalNetworks}
						& 		115	& 115	& 115	& 115		 & 115				& 115				&	115\\
\textit{lesmis} \cite{KnuthStanfordGraphBase}
						& 		49	& 49	& 49	& 49		& 49					& 49 				&	49\\
\textit{netscience} \cite{NewmanFindingCommnityStructureUsingTheEigenvectors}
						& 		&	108	& 107		& 107				& 107	 			& 108		& 108\\
\textit{zachary}	 \cite{ZacharyConflictAndFissionInSmallGroups}
						&		20 & 20		& 20			& 20		 	& 20		 			& 20 			&		20\\
\textit{celegansneural} \cite{WattsStrogatzCollectiveDynamicsSmallWorldNetworks}
						& 	280 &	280 &	280	& 270		 	& 267	 			& 271		& 280 \\
\textit{dolphins} \cite{LusseauBottlenoiseDolphins}
						& 		53	& 53			& 53		& 53		 	& 53			 		& 53		&	53\\
\textit{polbooks*}				&		105	& 105	& 105			& 104		 & 103	 			& 104	& 105	\\
\midrule
\multicolumn{7}{c}{\textit{Protein-protein interactions from UCLA database of interacting proteins} \cite{SalwinskiDatabaseOfInteractingProteins,XenariosDatabaseOfInteractingProteins2001,XenariosDatabaseOfInteractingProteins,XenariosDatabaseOfInteractingProteins2002}} \\
\textit{Celeg20160114CR}		&	6 & 6			& 6 & 6		 	& 6		 			& 6 & 6 \\
\textit{Dmela20160114CR}		&	 14 &	14	& 14		& 14		 		& 14		 			& 14 	& 14				\\
\textit{Ecoli20160114CR}			& 		& 242	& 242 & 207		 		& 205				& 211				&	\textbf{246}	\\
\textit{Hpylo20160114}			& 	& 291		& 291 & 239		 		& 235					& 241					&	\textbf{311}\\
\textit{Hsapi20160114HT}		&		64		& 64 & 64	& 63		 	& 63	 			& 64				&	64\\
\textit{Mmusc20160114CR}			& 	& 256	& 256	& 243		 	& 248					& 246				&	\textbf{278}\\
\midrule
\multicolumn{7}{c}{\textit{DIMACS graphs} \cite{JohnsonDimacs}} \\
\textit{anna}					&		& 79 & 79	& 76		& 77	 					& 77 					& 79\\
\textit{david}					&		72	& 72	& 72	& 71		 	& 70		 				& 71 			&		72	\\
\textit{homer}					&		&	 223 & 223	& 206		 			& 205					& 209			& \textbf{242}\\
\textit{huck}					& 		48 & 48	& 48	& 48		 & 48		 			& 48 					&48	\\
\bottomrule
\end{tabular}
\end{center}
\label{tab:resultsheuristic2}
* These solutions were found using our ILP formulation of the problem using the CBC solver with $1$ hour time limit per vertex. Not all of these solutions are therefore guaranteed to be optimal. It took days for some of the largest networks to find these solutions using CBC.
}
\end{table*}

Table 1 presents the detailed results obtained by the ANTH-LS algorithm in our experiments. The first column represents the name of the instance, followed by the optimum, if known, and the length of the longest cycle ever found previously for the corresponding instance. The following columns summarise how the ANTH-LS algorithm performed for the instance. The first column contains the length of the longest cycle found, the success rate in finding cycles of this length out of $10$ runs, the average number of generations and average CPU required.

In Table 2, we present a comparison of the results obtained by the ANTH-LS algorithm with other techniques to solve the problem. Values in the table represent the lengths of the longest cycles found by the corresponding methods. CBC represents the results obtained by the ILP-based approach using the branch-and-cut solver CBC, with a generous $1$-hour time limit per vertex. As some of these networks have hundreds of vertices, it took days to find some of the cycles reported for this method. MSLS algorithms represent multi-start local search repeatedly using DFS to construct long cycles, improved by LS-III or LS-IV. These algorithms used 100000 or 10000 restarts. The number of restarts and the use of LS-III or LS-IV are both indicated in the algorithm identifier.

The results that stand out most significantly are the previously unknown cycles that have been discovered by the ANTH-LS algorithm. These include the longest cycles so far for social network samples $gplus\_500$ and $pokec\_500$, as well as for the protein-protein interaction networks \textit{Ecoli20160114CR}, \linebreak \textit{Hpylo20160114}, \textit{Mmusc20160114CR} and coappearance network $homer$. It is possible that there are longer cycles in some of these networks. Hybridisation of the ANTH-LS algorithm, as well as its parallel or distributed variants could potentially help to discover these.

In addition, one can also observe that the ant-based algorithm is the first heuristic method to find optima for instances $gplus\_200$, $adjnoun$, \linebreak $celegansneural$, $polbooks$, and $david$. For these instances, the proven optima have previously been discovered in time-consuming procedures by the ILP-based approach and CBC \cite{CSI-0055}. The ANTH-LS algorithm was successful in reproducing these optima relatively quickly, indicating its high relevance for practical applications. For $netscience$ and $anna$, the proven optima are not yet known, even though both CBC and the ANTH-LS algorithm found a cycle of the same length.

\section{Conclusions}\label{sec:conclusions}

We proposed a new probabilistic ant-based heuristic (ANTH-LS) for finding long cycles in real-world complex networks. The algorithm uses the idea of extending depth-first search (DFS) construction of long cycles into a probabilistic scheme. In this strategy, the probability of traversing the edges, that have been present in long cycles previously found, is decreased. Conversely, the probability of traversing the other, new edges is reinforced. This idea serves as diversification mechanism, combined with an intensification local search subroutine based on four perturbation operators.

Our experimental results were presented for a set of diverse real-world networks, including social networks, network science data, as well as protein-protein interaction networks. These indicate that the ANTH-LS algorithm has found 6 previously unknown longest cycles for two social networks, three protein-protein interaction networks, as well as one coappearance network. In addition, the ANTH-LS algorithm is the first heuristic to quockly find previously known optima for a number of instances, previously only found by a computationally demading procedure based on integer linear programming.

Based on these results, we believe the ant-based algorithm is valuable especially for further real-world applications of long cycle identification. Such applications include graph drawing \cite{TamassiaHandbookOfGraphDrawing}, analysis of social networks \cite{CSI-0055}, layout algorithms for metabolic pathways \cite{BeckerGraphLayoutMetabolicPathways}, as well as analysis of utility distribution networks.
 
\small
\bibliographystyle{plain}
\bibliography{common}

\end{document}